\documentclass[11pt,aps,prc,superscriptaddress,showpacs]{revtex4}
\usepackage[dvips]{graphicx}
\usepackage{amsmath,amssymb}
\newcommand{\Nt}{\mbox{${\tilde N}_T$}}

\newcommand{\epsi}{\mbox{$\varepsilon$}}

\newcommand{\vbr}{\mbox{\boldmath $r$}}

\newcommand{\vq}{\mbox{\boldmath $q$}}
\newcommand{\vu}{\mbox{\boldmath $u$}}

\newcommand{\vp}{\mbox{\boldmath $p$}}

\newcommand{\vj}{\mbox{\boldmath $j$}}

\begin{document}

\title{Dipole Oscillations in Bose - Fermi Mixture 
\\in the Time-Dependent Grosspitaevskii and Vlasov equations}

\author{Tomoyuki~Maruyama}
\affiliation{
Institute for Nuclear Theory,
University of Washington,
Seattle, Washington 98195, USA}
\affiliation{College of Bioresource Sciences,
Nihon University,
Fujisawa 252-8510, Japan}
\affiliation{Advanced Science Research Center,
Japan Atomic Energy Research Institute, Tokai 319-1195, Japan}

\author{George F. Bertsch}
\affiliation{
Institute for Nuclear Theory,
University of Washington,
Seattle, Washington 98195, USA}

\date{\today}

\begin{abstract}
We study the dipole collective oscillations in bose-fermi mixtures
in a dynamical time-dependent approach, which is formulated with
the time-dependent Gross-Pitaevskii equation and the Vlasov equation.
While the bose gas oscillates monotonously, 
the fermion oscillation shows a beat and damping.
We find big differences in behaviors of fermion oscillation between 
the time-dependent approach and usual approaches such as 
the sum-rule approach. 
When the amplitude is not minimal,
the dipole oscillation of the fermi gas 
cannot be described with a simple center-of-mass motion.
\end{abstract}



\maketitle

\section{Introduction}
\label{intr}

Over the last several years, there have been significant progresses 
in the production of ultracold gases, which realize 
the Bose-Einstein condensates (BEC) \cite{nobel,Dalfovo,becth,Andersen}, 
two boson mixtures  \cite{B-B}, 
degenerate atomic Fermi gases \cite{ferG},
and Bose-Fermi (BF) mixing gases \cite{Schreck,BferM,Modugno}.   
In particular the BF mixtures attract physical interest 
as a typical example in which particles 
obeying different statistics are intermingled.
Using this system we have a very big opportunity to get
new various knowledge about many body systems 
because we can make a large variety of combinations of 
atomic species and control the atomic interactions 
using the Feshbach resonance \cite{Fesh}.
Theoretical studies of the BF mixtures have been done for static 
properties \cite{Molmer,Amoruso,MOSY,Bijlsma,Vichi,Vichi1}, 
for the phase diagram and phase
separation \cite{Nygaard,Yi,Viverit,Capuzzi02}, 
for stability  \cite{MSY1,Roth,Capuzzi03} and
for  collective excitations \cite{MSY,Minguzzi,zeros,Yip,sogo,miyakawa,Liu,tomoBF}. 

Above all phenomena the spectrum of the collective excitations is
an important diagnostic signal for these systems.
Such oscillations are common to a variety of many-particle systems
and are often sensitive to the interaction and the structure of the
ground state and the excited states.  
Theoretically collective motions are usually studied with 
the random phase approximation (RPA) \cite{zeros,sogo} or
its approximate methods such as the sum-rule \cite{miyakawa}
and the scaling \cite{Liu,scal,scalDP} approaches.
The first author (T.M.) and his collaborators \cite{tomoBF} 
studied monopole oscillations of BF mixtures
by calculating the time evolution of these oscillations with 
the time-dependent Gross-Pitaevskii (TDGP) equation and the Vlasov equation.
This dynamical approach showed 
different behaviors 
of the oscillation from RPA \cite{sogo}, such as
rapid damping at zero temperature.

The RPA can treat only  states
with one-particle one-hole excitation 
and only describe minimal vibrations  around ground states;
for example it is shown in Ref.\cite{sogo} that the radial variation 
is only about 0.05 \% in RPA, while the amplitudes of actual experiments 
can be as large as 10 \% \cite{ColEx}.

For the single boson system 
the RPA \cite{ColBTh} explained experimental results 
on frequencies of collective motions well \cite{ColEx}.
In this system  
all bosons occupy one single particle state 
at zero temperature, and their collective motions are simple;
damping does not appear at very low temperature 
in experiments \cite{chevy} and in a theoretical work with
the time-dependent density matrix theory \cite{tohyama}.

On the other hand fermions occupy many single particle states even 
at zero temperature.
When the amplitude is about 10 \% of RMSR, 
the monopole states have as large excited energy as ten to one-hundred times 
of the one-particle and one-hole excitation energy.
Their collective oscillations with a large amplitude 
are multiparticle and multihole states and
include various mode with different frequencies.
Indeed
an oscillation of  population difference in two-component fermi gas has
shown damping due to a multimode dephasing \cite{Poetting}.
In the BF mixtures, especially,
fermions occupy a larger region than condensed bosons,
and have different potentials between inside and outside 
of the boson occupation region.
Then their motions are not harmonic, and then the fermion oscillation
makes damping \cite{tomoBF}.
In order to study the collective oscillation in BF mixtures, hence, 
we need to calculate time evolution of the system using
a time dependent dynamical approach.

Thus the BF mixing gases show new dynamical properties different from
those in other finite many-body system such as nuclei.
In order to have more information on the dynamical properties,
we need to investigate other kinds of multipole motions.
In this paper we study the dipole oscillations
in bose-fermi mixtures as the next step by solving
the time evolution of the condensed boson wave function
and the fermion phase-space distribution function
with the TDGP equations and the Vlasov equations,
respectively. 

In the next section we explain our transport model
to calculate the time evolution of the system. 
In Sec. III we show the calculational results for 
the dipole oscillation in the BF mixture, 
and discuss  their properties.
Then we summarize our work in Sec. IV.

\section{Time Evolution Equations}
\label{TevEq}

Here we briefly explain our approach.
In this work we consider a dilute boson and one-component-fermion 
coexistent gases at zero temperature with the axial symmetry 
with respect to the $z$-axis. 
We assume only zero-range interaction between atoms, and
there is no fermion-fermion interaction in the system. 
The hamiltonian is written as
\begin{eqnarray}
{\tilde H} &=&  \int d^3 q ~\left[
- \frac{\hbar^2}{2 M_B} {\tilde \phi}^{\dagger}(\vq) \nabla^2_q
{\tilde \phi}(\vq)
+ \frac{1}{2} M_B \Omega_B^2 (\vq_T^2 + \kappa_L^2 q_L^2) {\tilde \phi}^{\dagger}(\vq)
{\tilde \phi}(\vq) 
+ \frac{2 \pi \hbar^2 a_{BB}}{M_B} \{ {\tilde \phi}^{\dagger} (\vq) {\tilde \phi} (\vq) \}^2
\right.
\nonumber \\
&  &  - \frac{\hbar^2}{2M_f} {\tilde \psi}^{\dagger} (\vq) \nabla_q^2 {\tilde \psi} (\vq)
+ \frac{1}{2} M_f  \Omega_F^2 
(\vq_T^2 + \kappa_L^2 q_L^2) {\tilde \psi}^{\dagger}(\vq) {\tilde \psi} (\vq)
\nonumber \\
& & \left. + \frac{2 \pi \hbar^2 a_{BF}}{M_r}
{\tilde \phi}^{\dagger} (\vq) {\tilde \phi} (\vq) 
{\tilde \psi}^{\dagger} (\vq) {\tilde \psi} (\vq) \right],
\end{eqnarray}
where
${\tilde \phi}$ and ${\tilde \psi}$ are boson and fermion fields, respectively, 
$M_B$ and $M_F$ are the boson and fermion masses,
$M_r \equiv M_B M_F/(M_B + M_F)$ is the reduced boson-fermion mass,
$\Omega_B$ and $\Omega_F$ are the transverse trapped frequencies 
of the boson and the fermion,
$a_{BB}$ and $a_{BF}$ are the $s$-wave scattering lengths 
between two bosons and between boson and fermion, respectively.
In addition the positional coordinate is described 
as $\vq \equiv (\vq_T, q_L)$,
and $\kappa_L$ is the ratio of the longitudinal trapped frequency to 
the transverse trapped frequency.

In this formulation we can change all variables to dimensionless ones
without losing generality as follows.
We normalized the spatial coordinate $\vq$ to the dimensionless
coordinates $\vbr$ as $\vbr \equiv (\vbr_T, z) \equiv \vq /R_B$ 
with $R_B = (\hbar / M_B \Omega_B)^{1/2}$.
According to this normalization the boson and fermion fields are also scaled as
$\phi (\psi) = R_B^{-1/3} {\tilde \phi} ( {\tilde \psi})$.
By dividing the above hamiltonian ${\bar H}$ by $\hbar \Omega_B$, then,
we can define the dimensionless hamiltonian $H \equiv {\tilde H} / \hbar \Omega_B$ as
\begin{eqnarray}
H &=&  \int d^3 r ~\left[
- \frac{1}{2} \phi^{\dagger}(\vbr) \nabla^2_r \phi(\vbr)
+ \frac{1}{2} (\vbr_T^2 + \kappa_L^2 z^2) \phi^{\dagger}(\vbr) \phi(\vbr) 
+ \frac{g_{BB}}{2} \{ \phi^{\dagger} (\vbr) \phi (\vbr) \}^2  \right.
\nonumber \\
 & &
 - \frac{1}{2m_f} \psi^{\dagger} (\vbr) \nabla^2_r \psi (\vbr)
+ \frac{1}{2} m_f \omega_f^2 
(\vbr_T^2 + \kappa_L^2 z^2) \psi^{\dagger}(\vbr) \psi (\vbr) 
\nonumber \\
&& \left.  + h_{BF} 
\phi^{\dagger} (\vbr) \phi (\vbr) 
\psi^{\dagger} (\vbr) \psi (\vbr) \right],
\end{eqnarray}
where $m_f \equiv M_F/M_B$, $\omega_f \equiv \Omega_F/\Omega_B$, and
$g_{BB} \equiv 8 \pi \hbar a_{BB} R_B^{-1}$
and  $h_{BF} \equiv 4 \pi \hbar m_f a_{BF} (1+m_f)^{-1} R_B^{-1}$.

In this work we consider the zero-temperature system,
including $N_b$ bosons and $N_f$ fermions,
so that the total wave function is written as
\begin{equation}
\Phi (\tau) = \{\prod_{i=1}^{N_b} \phi_c (\vbr_i) \} \Psi_f[\psi_n] , 
\end{equation}
where $\phi_c$ is a wave function of the condensed boson and
$\Psi_f$ is a Slater determinant of fermions 
with single particle wave functions, $\psi_n$.
  
The time evolution of the wave functions are obtained from the variational
condition that
\begin{equation}
\delta \int d \tau 
<\Phi (\tau) |\{ i\frac{\partial}{\partial \tau} - H \} |\Phi (\tau) > = 0 .
\label{vari}
\end{equation}
From this condition we derive coupled equations
of the TDGP and TDHF equations as follows.
\begin{eqnarray}
i \frac{\partial}{\partial \tau}  \phi_c (\vbr, \tau) &=&
\left\{ - \frac{1}{2} \nabla_r^2 + U_B (\vbr) \right\} ~ \phi_c (\vbr, \tau) ,
\label{TDGP}
\\
i \frac{\partial}{\partial \tau}  \psi_n (\vbr, \tau) &=&
\left\{ - \frac{1}{2 m_f} \nabla_r^2 + U_F (\vbr) \right\} ~\psi_n (\vbr, \tau)
\label{TDHF}
\end{eqnarray} 
with 
\begin{eqnarray}
U_B (\vbr) &=& \frac{1}{2} (\vbr_T^2 + \kappa_L^2 z^2)
 + g_{BB} \rho_B (\vbr) + h_{BF} \rho_F (\vbr) ,
\label{uB}
\\
U_F (\vbr) &=& \frac{1}{2} m_f \omega^2_f (\vbr_T^2 + \kappa_L^2 z^2)
 + h_{BF} \rho_B (\vbr) ,
\label{uF}
\end{eqnarray} 
where $\rho_B$ and $\rho_F$ are boson and fermion densities which are given as
\begin{eqnarray}
\rho_B (\vbr) &=& N_b |\phi_c (\vbr)|^2 ,
\label{rhoB}
\\
\rho_F (\vbr) &=& \sum^{occ}_{n} |\psi_n (\vbr)|^2 .
\label{rhoF}
\end{eqnarray} 

The number of fermion states are usually too large 
to solve the above TDHF equations directly so instead 
one uses the semi-classical approach.
In the semi-classical limit ($\hbar \rightarrow 0$)
the TDHF equation is equivalent to  
the following Vlasov equation \cite{KB}:
\begin{equation}
\frac{d}{d \tau} f(\vbr,\vp;\tau) =
\left\{ \frac{\partial}{\partial \tau} + \frac{\vp}{m_f}{\nabla_r} -
 [\nabla_r U_F(\vbr)][\nabla_p] \right\} f(\vbr,\vp;\tau) = 0 ,
\label{Vlasov}
\end{equation}
where $f(\vbr,\vp;\tau)$ is the fermion phase-space distribution function 
defined  as
\begin{equation}
f(\vbr,\vp,\tau) = \int {d^3 u} <\Phi|
\psi(\vbr+\frac{1}{2}{\vu},\tau)
\psi^{\dagger}(\vbr-\frac{1}{2}{\vu},\tau) |\Phi>
 e^{-i \vp \vu } .
\end{equation}

As an actual numerical method we introduce the test particle method \cite{TP}
to solve the Vlasov equation (\ref{Vlasov}) and 
describe the fermion phase-space distribution function as 
\begin{equation}
f(\vbr,\vp,\tau) = \frac{(2 \pi)^3}{\Nt} 
\sum_{i=1}^{{\tilde N}_T N_f} \delta\{\vbr-\vbr_i(\tau)\} \delta\{\vp-\vp_i(\tau)\} .
\label{TP-Wig}
\end{equation}
where {\Nt} is the number of test-particles per fermion.

By substituting Eq.(\ref{TP-Wig}) into Eq.(\ref{Vlasov}),
we can obtain  the following equations of motion for test-particles as
\begin{eqnarray}
\frac{d}{d \tau} \vbr_i (\tau) &=& \frac{\vp_i}{m_f},
\label{eqM1} \\
\frac{d}{d \tau} \vp_i (\tau) &=& - \nabla_r U_F(\vbr).
\label{eqM2}
\end{eqnarray}

Thus we obtain the time evolutions
of the condensed boson wave function and the fermion phase space
distribution function
by solving eqs.(\ref{TDGP}), (\ref{eqM1}) and  (\ref{eqM2}).

\section{Results and Discussions}
\label{ResD}

In this section we show the results of our calculations  
on the dipole oscillations in the BF mixture.
We deal with the system $^{87}$Rb~-$^{40}$K, where
the number of the bosons ($^{87}$Rb) and the fermions ($^{40}$K)
are $N_b = 10000$ and $N_f = 1000$, respectively, and 
the fermion mass normalized by boson mass  is $m_f=40/87 \approx 0.46$. 
We assume the spherical trap ($\kappa_L = 1$) and
take the fermion trapped frequency to be 
$\omega_f = 1/\sqrt{m_f} \approx 1.48$.
The boson-boson interaction parameter  $g_{BB}$ is fixed
to be $g_{BB} = 1.34 \times 10^{-2}$ \cite{gBB-C},  which corresponds to 
$a_{BB} = 4.22$(nm), while 
the BF interaction parameter $h_{BF}$ is varied.
In the numerical calculation we take 
the number of the test-particles per fermion to be ${\tilde N} = 100$
and solve the time evolutions
by using the second order predictor-corrector method.

\subsection{Ground State}
\label{GRD}

The sum-rule approach \cite{miyakawa} showed that 
the dipole oscillation frequency is sensitive to 
the density distributions.
In this subsection, then, we see the boson and fermion density 
distribution from 
the point of view of the boson-fermion coupling dependence.

In the ground state the 
the wave function of the condensed boson, $\phi_c$ is defined as a
solution of the following Gross-Pitaevskii equation:
\begin{eqnarray}
\left\{ - \frac{1}{2} \nabla_r^2 
+ \frac{1}{2} (\vbr_T^2 + \kappa_L^2 z^2)
 + g_{BB} \rho_B (\vbr) + h_{BF} \rho_F (\vbr) \right\} ~ 
\phi^{(g)}_c (\vbr)  = \mu_b  \phi_c^{(g)} (\vbr) ,
\label{GPgr}
\end{eqnarray}
where $\mu_b$ is the boson chemical potential.
In the ground state
the fermion phase-space distribution function is given by
the Thomas-Fermi (TF) approximation
\begin{equation}   
f (\vbr,\vp ) = \theta[ \mu_f - \epsi(\vbr,\vp) ] ,
\end{equation}
with
\begin{equation}
 \epsi(\vbr, \vp) = \frac{1}{2m_f} \vp^2 + U_F (\vbr) ,
\end{equation}
where $\mu_f$ is the fermion chemical potential.
In this TF approximation the fermion density $\rho_F$ is 
obtained as the solution of  the following equation:
\begin{equation}
\frac{1}{2 m_f} \left\{6 \pi^2 \rho_F (\vbr) \right\}^{2/3}
+ \frac{1}{2} m_f \omega^2_f (\vbr_T^2 + \kappa_L^2 z^2)
 + h_{BF} \rho_B (\vbr) = \mu_f .
\label{TFf}
\end{equation}
Here we iterate solving the boson wave function with Eq.(\ref{GPgr}) 
and searching the fermi energy $\mu_f$ in Eq.(\ref{TFf}) 
to give the correct fermion number.

\begin{figure}[ht]
\hspace*{0cm}
{\includegraphics[scale=0.8]{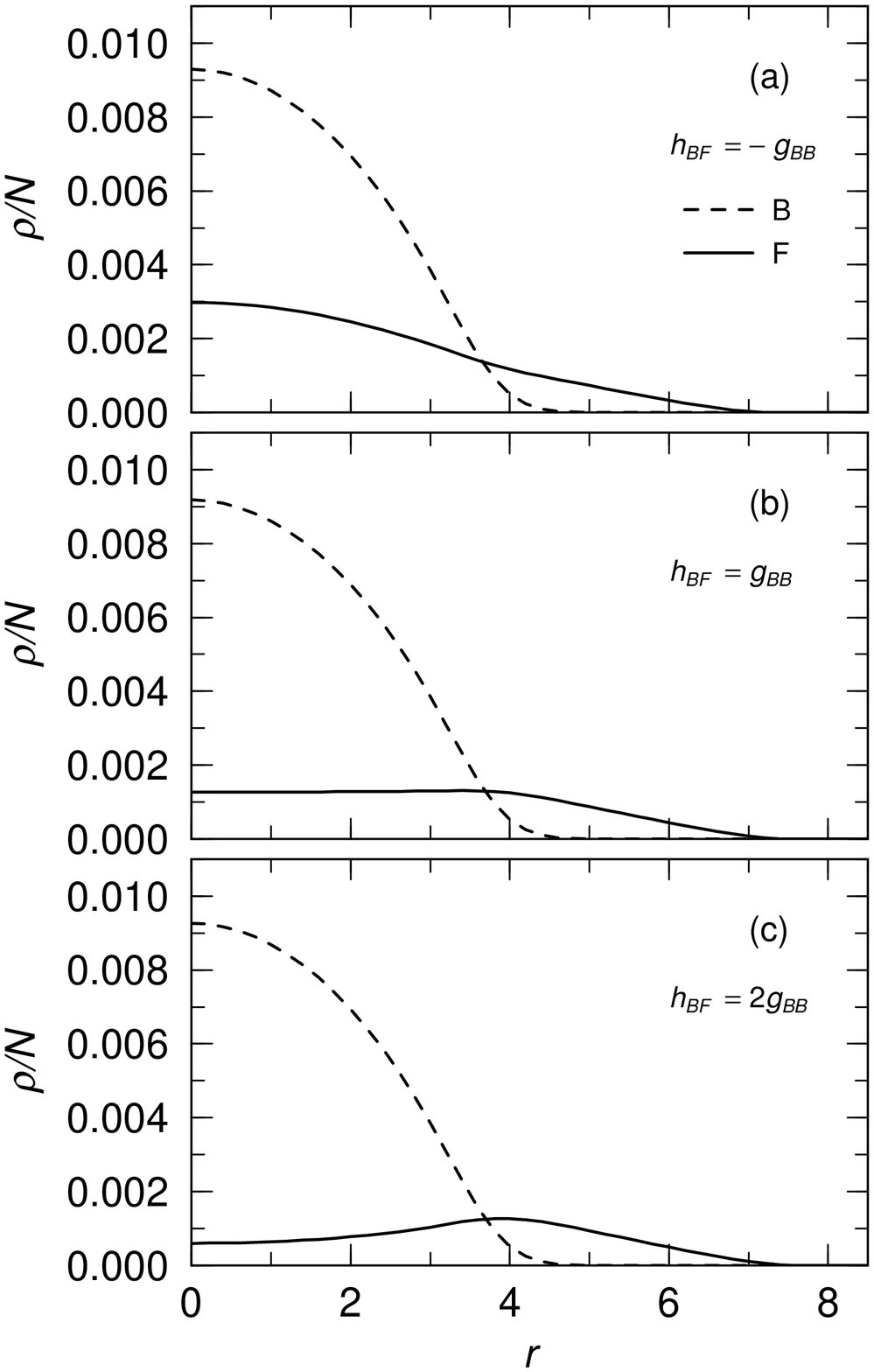}}
\caption{\small
The density distribution of the BF mixing gas
with the boson-fermion coupling $h_{BF} = - g_{BB}$ (a),
$h_{BF} = g_{BB}$ (b), and $h_{BF} = g_{BB}$ (c).
The dashed and solid lines represent the density of
the boson and fermi gases, respectively.}
\label{grdRH}
\end{figure}

In Fig.~\ref{grdRH} we show the density distribution of 
the boson and fermi gases with the boson-fermion coupling
$h_{BF} = - g_{BB}$ (a), $h_{BF} = g_{BB}$ (b)  
and $h_{BF} = 2 g_{BB}$ (c).
The solid and dashed lines represent the results of
boson and fermi gases, respectively.
The density distribution of the bose gas is negligibly changed 
while the fermion density distribution in the boson occupation region 
varies as the boson-fermion coupling increases.
The fermion density distribution 
is central peaked when $h_{BF} = - g_{BB}$, 
flat when $h_{BF} = g_{BB}$,
and surface peaked when $h_{BF} = 2g_{BB}$.

This boson-fermion coupling dependence of the fermion density
distribution 
can be easily explained as follows.
When the boson number, $N_b$, is very large,
the boson density distribution can also be given by the TF approximation
as
\begin{equation}
\rho_B (\vbr) = \frac{1}{g_{BB}}
\left[ \mu_B - \frac{1}{2} (\vbr_T^2 + \kappa_L^2 z^2) 
- h_{BF} \rho_F (\vbr) \right] .
\label{TFb}
\end{equation}
Note that the boson density in Eq.(\ref{TFb}) is defined within a
region, and it is zero outside of that region.

Substituting Eq.(\ref{TFb}) into Eq.(\ref{TFf}), we can get
\begin{equation}
\frac{1}{2 m_f} \left(6 \pi^2 \rho_F \right)^{2/3}
 - \frac{h^2_{BF}}{g_{BB}} \rho_F
= \mu_F - \frac{h_{BF}}{g_{BB}} \mu_B
- \frac{1}{2}
\left( m_f \omega^2_f - \frac{h_{BF}}{g_{BB}} \right) 
\xi^2 ,
\label{TFf2}
\end{equation}
where $\xi^2 \equiv \vbr_T^2 + \kappa_L^2 z^2$.
From this equation we can easily know that 
$\rho_F = const.$  when $h_{BF}/g_{BB}=1$ and $\rho_B > 0$.
Furthermore the derivative of the fermion density, $\rho_F$, with
respect to $\xi$ is given as
\begin{equation}
\left( \frac{4\pi^4}{3 m_f^3}\right)^{\frac{1}{3}}
\left\{  \rho_F^{-1/3}
 - \frac{3m_f^3 h^6_{BF}}{4 \pi^4 g^3_{BB}}  \right\}  
\frac{1}{\xi} \frac{\partial \rho_F }{ \partial \xi }
= - \left( m_f \omega^2_f - \frac{h_{BF}}{g_{BB}} \right) .
\label{TFfDr}
\end{equation}
The stability condition \cite{Yi}, that 
$(\partial \mu_b/\partial \rho_B)(\partial \mu_f/\partial \rho_F) -
(\partial \mu_b/\partial \rho_F)(\partial \mu_f/\partial \rho_B)  > 0$,
restricts the value of the fermion density as
 $\rho_F < 4 \pi^4 g_{BB}^3/3 m_f^3 h_{BF}^6 $ .
In the boson occupation region, therefore,
the derivative of the fermion density is negative,  
$\partial \rho_F / \partial \xi < 0$, when 
$h_{BF}/ g_{BB} > m_f \omega_f^2$,
and positive,  $\partial \rho_F / \partial \xi > 0$, 
when $h_{BF}/ g_{BB} > m_f \omega_f^2$.

As mentioned before, we take the parameters to be  
$m_f \omega_f^2 = 1$ and $\kappa_L=1$  in the present calculation.
Thus the TF approximation can explain the relation between the boson-fermion coupling 
and the fermion density distributions.

\subsection{Dipole Oscillation}
\label{DP}

In this subsection we show our actual results of the numerical simulations on 
the dipole oscillations.
Here we define the center of mass (CM) position
on $z$-coordinates  for bosons and fermions as $z_B$ and $z_F$, respectively. 
We will discuss the oscillation behavior by examining 
 the time-dependence of $z_B$ and $z_F$.

In actual simulations we boost the condensed boson wave function and 
the fermion test-particles at the starting time, $\tau= 0$,  
in the following way:
\begin{equation}
\phi_c(\vbr, \tau=0) = e^{i \lambda_B z} \phi_c^{(g)} (\vbr) ,
\noindent
\label{bin}
\end{equation}
\begin{equation}
p_z(i) = p_z^{(g)}(i) + m_f \omega_f \lambda_F
\label{fin}
\end{equation}
with  the boost parameters $\lambda_B$ and $\lambda_F$,
where the superscript $(g)$ represents the wave function and 
the coordinates of the ground state.
These transformations give the current density of boson,
$\vj_B(\vbr)$, and fermion, $\vj_F(\vbr)$, as
\begin{eqnarray}
\vj_B(\vbr,\tau=0) &=& \rho_B^{(g)} (\vbr) \lambda_B {\hat z} ,
\\ 
\vj_F(\vbr,\tau=0) &=& \rho_F^{(g)} (\vbr) \omega_f \lambda_F {\hat z} .
\end{eqnarray}
If $h_{BF} = 0$, the time dependences of $z_{B,F}$ become
$z_B = \lambda_B \sin (\tau)$ and $z_F = \lambda_F \sin (\omega_f \tau)$;
the boost parameters $\lambda_B$ and $\lambda_F$ correspond
to the initial amplitudes of $z_B$ and $z_F$, respectively.

\begin{figure}[ht]
\hspace*{0cm}
{\includegraphics[scale=0.85]{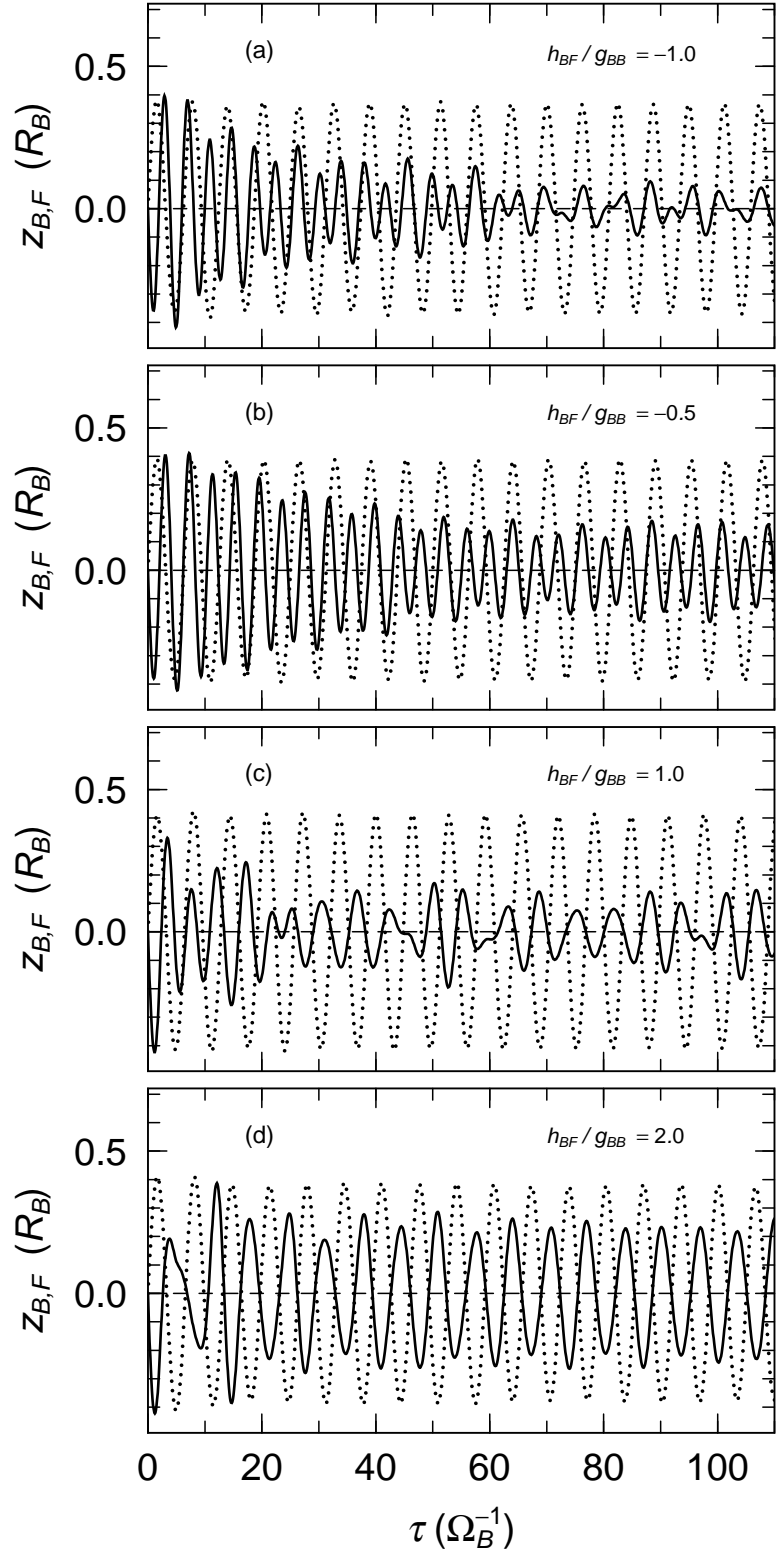}}
\caption{\small
Time evolution of the center-of-mass position 
of the boson (dotted lines) and fermion (solid lines)
with the boson-fermion coupling $h_{BF} = - g_{BB}$ (a),
$h_{BF} = - 0.5 g_{BB}$ (b),
$h_{BF} = g_{BB}$ (c) and $h_{BF} = 2 g_{BB}$ (d).
The initial condition is taken to be $\lambda_B = 0.4$ 
and $\lambda_F = - 0.4$.}\
\label{dplRbK}
\end{figure}

In Fig.~\ref{dplRbK} we show the time-dependence of the 
$z_{B,F}$ when the boson-fermion coupling
$h_{BF} = - g_{BB}$ (a), $h_{BF} = - 0.5g_{BB}$ (b),
$h_{BF} = g_{BB}$ (c) and $h_{BF} = 2 g_{BB}$ (d).
The dashed and solid lines represent the results of $z_B$ and $z_F$, respectively.
In all calculations we choose the out-of-phase at the beginning  
between the boson and fermion oscillations
by taking the initial condition to be $\lambda_B=0.4$ and $\lambda_F = -0.4$.

While the boson oscillations are monotonous,
the fermion oscillations have damping in the early time stage.
When $h_{BF}=\pm g_{BB}$,
the amplitude of the fermion oscillation is about 0.4 
at the beginning, but it decreases and becomes
about 0.1 $-$ 0.2 after the damping ($\tau \gtrsim 60$).
This damping becomes slower when the coupling is weaker,
$h_{BF} =- 0.5g_{BB}$ (Fig.~\ref{dplRbK}b) .
Furthermore, we see that after the damping 
$z_B$ and $z_F$ oscillate with the same period, 
and their relative phase becomes in-phase when $h_{BF} < 0$ 
and out-of-phase when $h_{BF} > 0$.
Note that, when $h_{BF} = 2g_{BB}$, the damping of $z_F$ 
is not clearly seen; 
the damping is too fast, and the period of the oscillation becomes
the same with that of $z_B$ in the early time stage.

\begin{figure}[ht]
{\includegraphics[angle=270,scale=0.85]{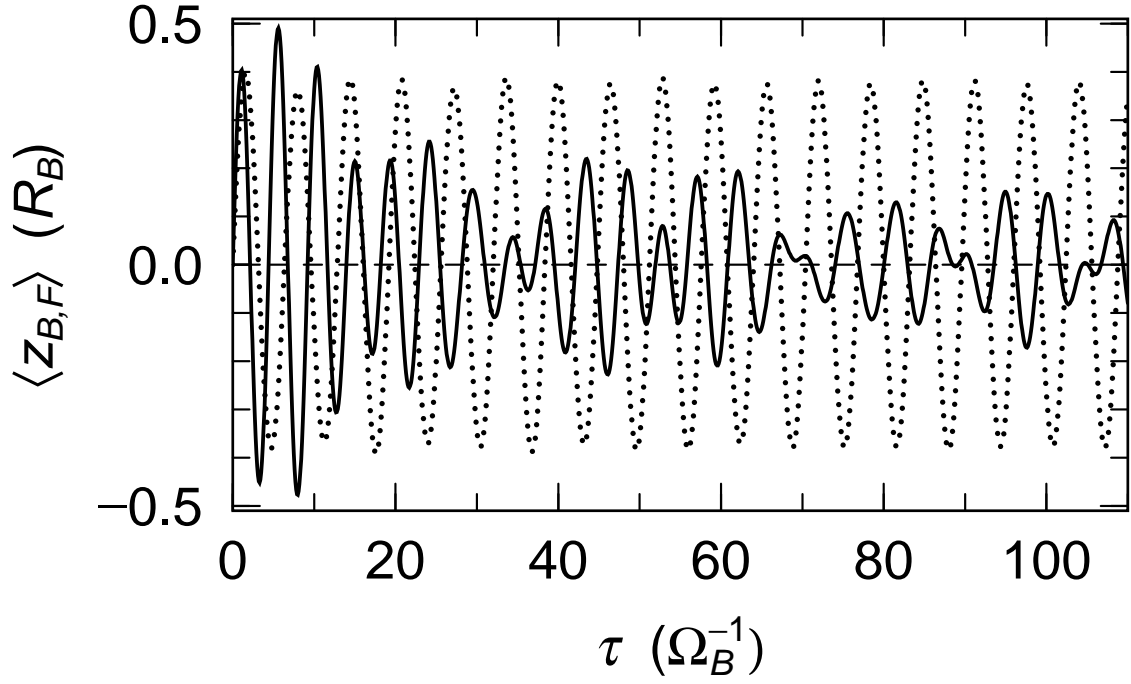}}
\caption{\small
Same as Fig.~\ref{dplRbK}c, but using the in-phase initial condition
 ($\lambda_B = \lambda_F = 0.4$) at the beginning.
}
\label{dpP1in}
\end{figure}

\begin{figure}[ht]
\hspace*{0cm}
{\includegraphics[scale=0.8]{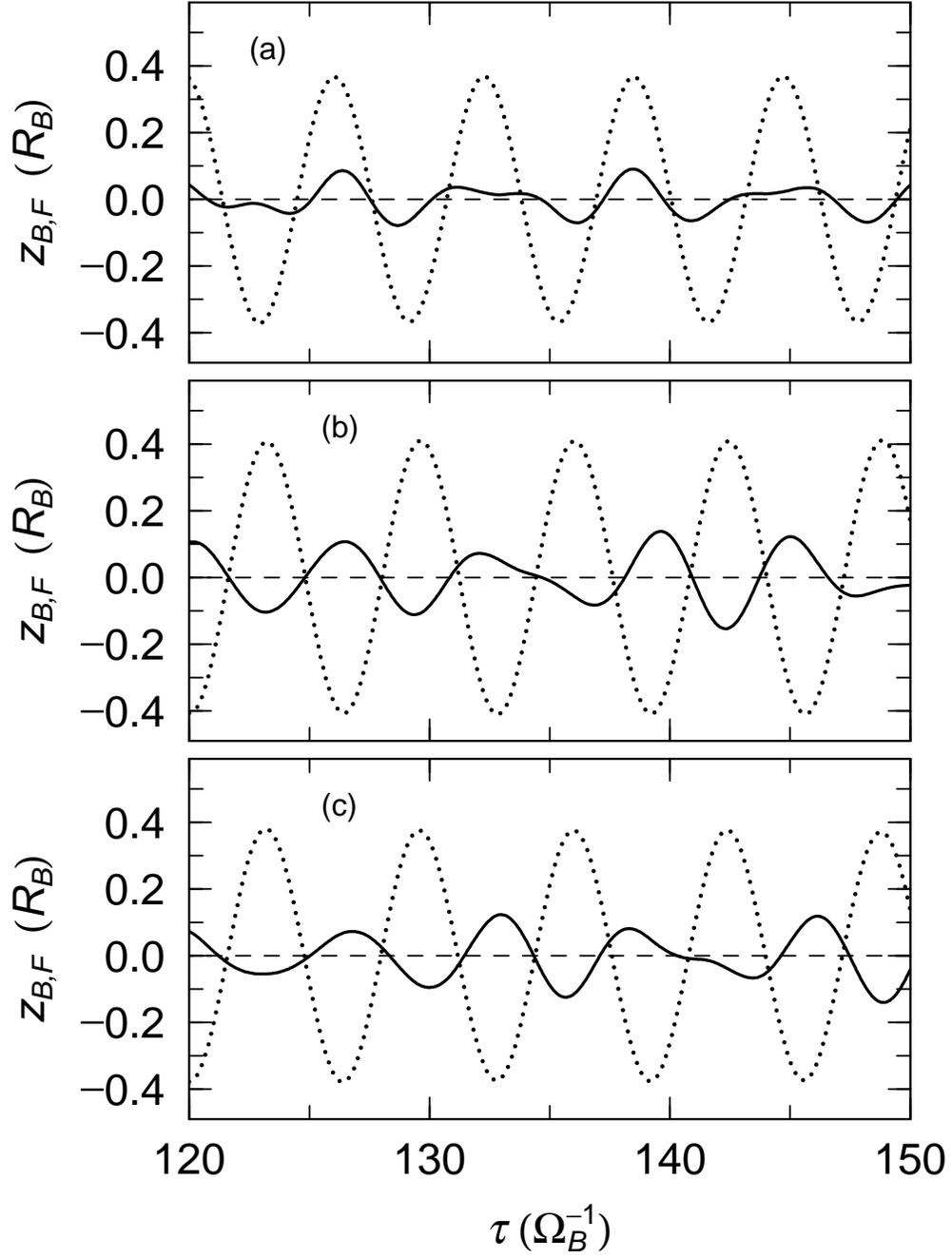}}
\caption{\small
Time evolution of $z_B$ (dashed lines) and $z_F$ (sold lines)
with the boson-fermion coupling $h_{BF} = - g_{BB}$ (a), 
$h_{BF} = g_{BB}$ (b) and  $h_{BF} = g_{BB}$ (c). 
In the first and second panels the oscillation is started with
the out-of-phase at the beginning, and the last ones with
the in-phase. }
\label{dpFi}
\end{figure}

In order to confirm the above comment about the relative phase 
after the damping, 
we calculate the dipole oscillation with $h_{BF}=g_{BB}$, 
which is started with in-phase at the beginning
($\lambda_B=\lambda_F = 0.4$).
The results are shown in Fig.~\ref{dpP1in}.
The relative phase between the boson and fermion oscillations 
becomes out-of-phase in the later time stage
after the damping.
In order to clarify it, we plot the same quantities shown in
Fig.~\ref{dplRbK}a, Fig.~\ref{dplRbK}d and Fig.~\ref{dpP1in}
in later time, $120 < \tau < 150$, in Fig.~\ref{dpFi}.
We can see that $z_F$ oscillates almost with the same period
of $z_B$.

\begin{figure}[ht]
\hspace*{-1.0cm}
\includegraphics[scale=0.65,angle=270]{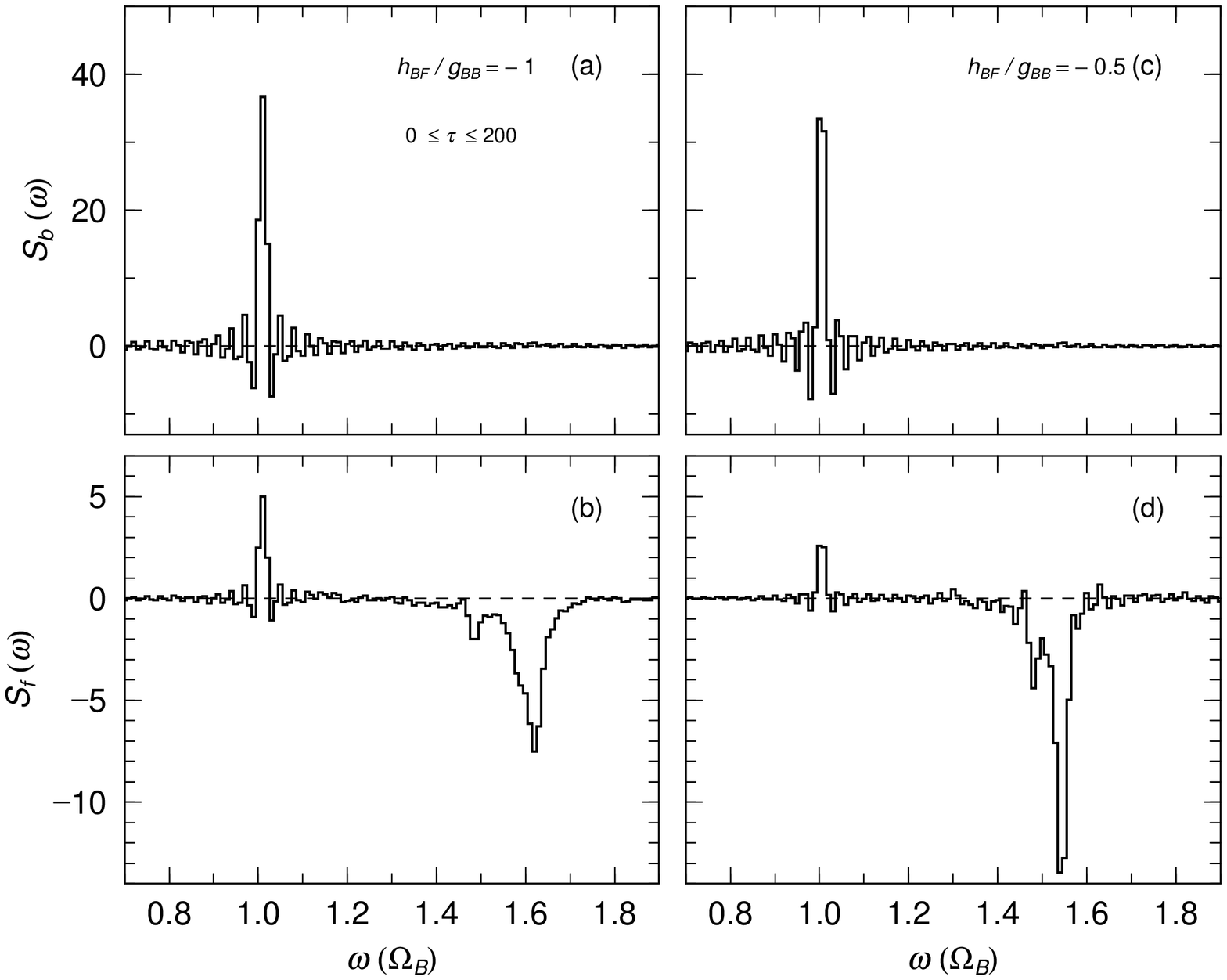}
\caption
{\small 
Strength functions of the boson oscillation in upper panels (a,c) 
and fermion oscillation in lower panels (b,d)
with $h_{BF}=-g_{BB}$ (a,b) and with $h_{BF}=-0.5 g_{BB}$ (c,d).
}
\label{spectF}
\end{figure}

\begin{figure}[ht]
\hspace*{-1.0cm}
\includegraphics[scale=0.65,angle=270]{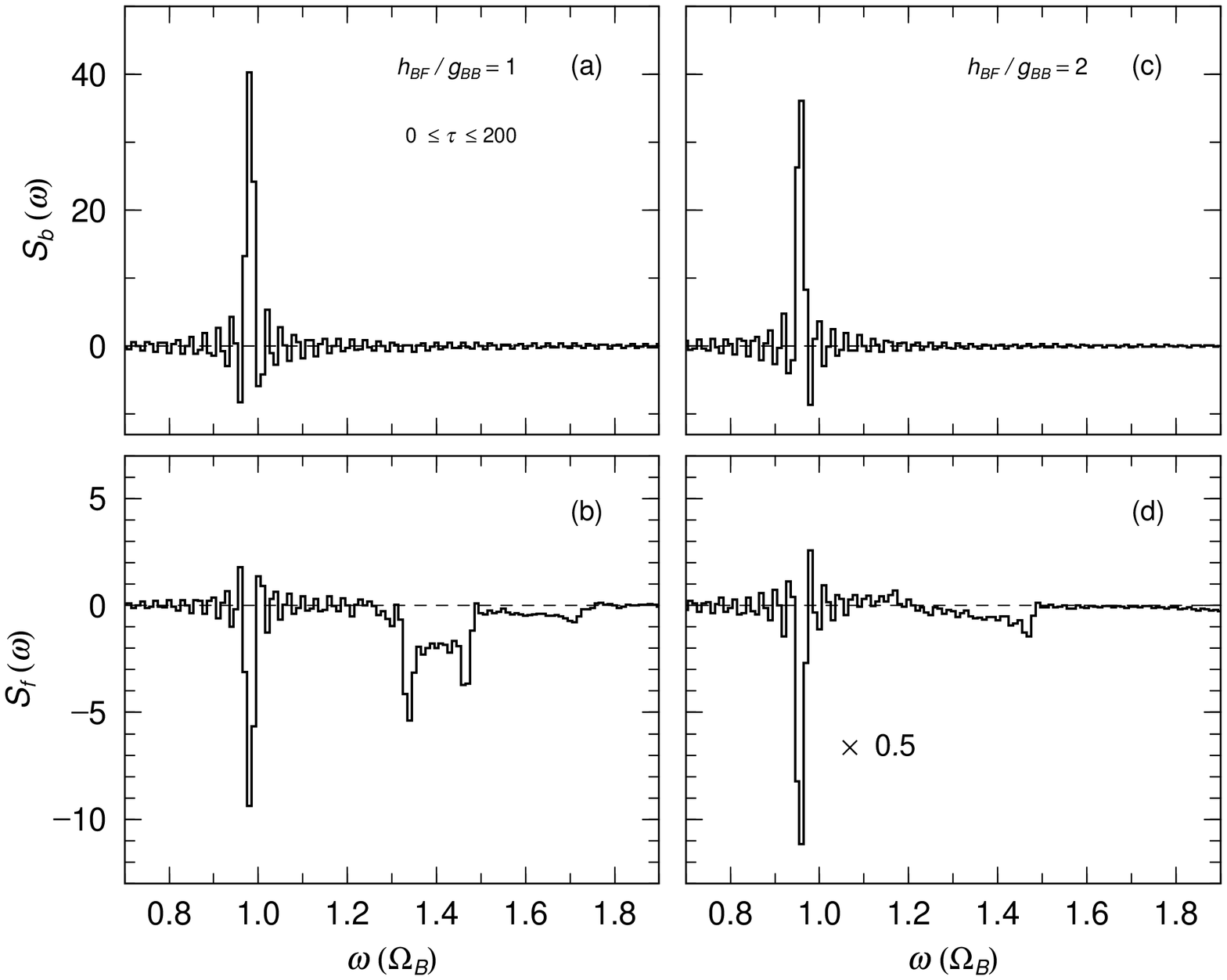}
\caption
{\small 
Strength functions of the boson oscillation 
in upper panels (a,c) and
fermion oscillation in lower panels (b,d)
with $h_{BF}= g_{BB}$ (a,b) and with $h_{BF}=2g_{BB}$ (c,d).
}
\label{spectH}
\end{figure}

As seen in Figs.~\ref{dplRbK}$-$\ref{dpFi}, 
the behaviors of the fermion oscillations are not so 
simple and imply that the fermion oscillation includes various modes.
In order to inspect these properties more, 
we calculate the strength function defined as the Fourier transformation of
$z_{B,F}$:
\begin{equation}
S_{B,F}(\omega) = 
\int^{t_f}_{t_i} d \tau z_{B,F} (\tau) \sin (\omega \tau) .
\label{spectE}
\end{equation}
In this work we fix that $t_i=0$ and $t_f=200$ ($\Omega_B^{-1}$). 
We show the strength 
functions of the dipole oscillations with $h_{BF}=-g_{BB}$ 
for the bosons (a) and the fermions (b) 
and with $h_{BF}= -0.5 g_{BB}$ for the bosons (c) and the fermions (d) 
in Fig.~\ref{spectF}, 
and those with $h_{BF}= g_{BB}$ for the bosons (a) and the fermions (b) and 
$h_{BF}= 2 g_{BB}$ for the bosons (c) and the fermions (d) in Fig.~\ref{spectH}.

First we note that  the boson strength functions have only one sharp peak, 
which is consistent with the monotonous behavior of the boson
oscillations.
Hence the frequency of the peak position can be considered
to be the intrinsic frequency of the boson oscillation;
we define $\omega_D^b$ as this frequency.
Second the fermion strength has three peaks.
One peak appears at $\omega = \omega_D^b$, 
and another peak appears at the trapped frequency,
$\omega = \omega_f \approx 1.47$ in all cases,
whereas the position of the other peak depends on the coupling constant.
These three peaks must correspond to certain modes.
For convenience we refer to these three modes as
mode-1, mode-2 and mode-3 in order, and 
define $\omega_D^f$ as the frequency of the mode-3.

\begin{figure}[ht]
\hspace*{-1.0cm}
\includegraphics[scale=0.7]{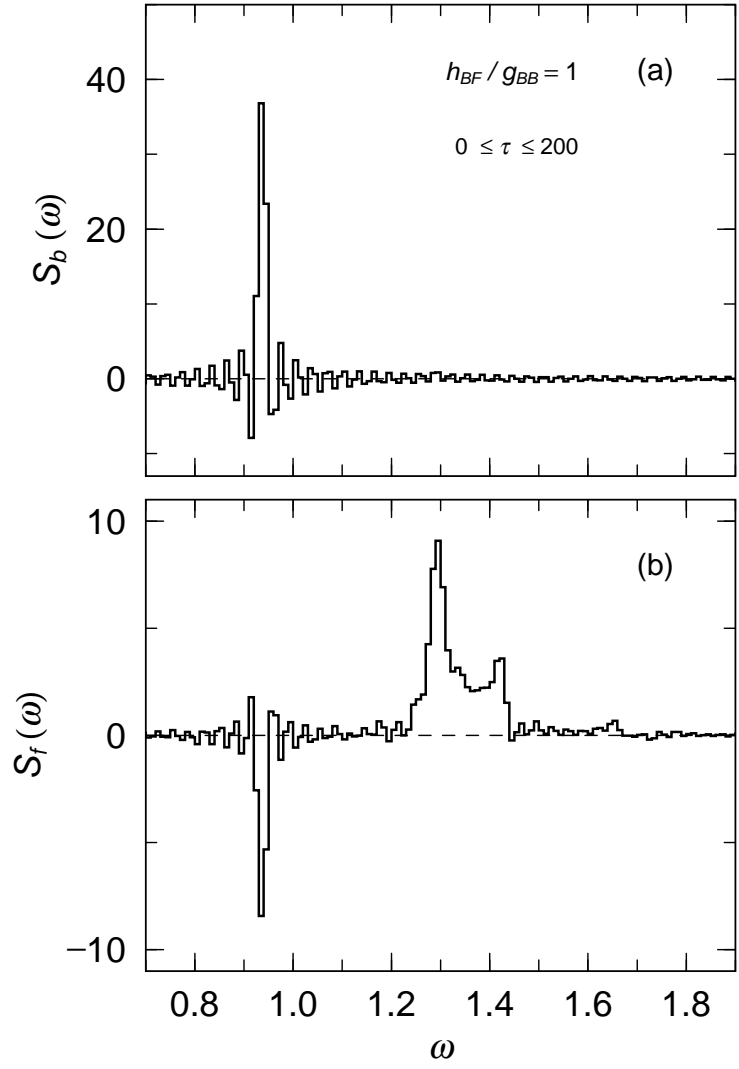}
\caption
{\small 
The strength function of the fermion dipole oscillation 
with $h_{BF}=g_{BB}$, 
but using the in-phase initial condition
 ($\lambda_B = \lambda_F = 0.4$).
}
\label{specP1in}
\end{figure}

The sign of the fermion strength function
at $\omega=\omega_D^b$ (mode-1) is plus when $h_{BF} > 0$ and 
minus when $h_{BF} < 0$, while
the signs of the strengths at $\omega= \omega_f$ (mode-2)
and  $\omega= \omega_D^f$ (mode-3) are minus in all oscillations.
In Fig.~\ref{specP1in}, furthermore, we plot the strength functions
of the bosons (a) and the fermions (b) with $h_{BF} = g_{BB}$ using
the in-phase initial condition ($\lambda_B = \lambda_F = 0.4$).
We see that the signs of the strength functions at $\omega= \omega_f$ 
and  $\omega= \omega_D^f$ become plus. 

In this initial condition the sign of the boson strength function
is always plus.
The sign of the strength exhibits the relative phase between
each mode and the boson oscillation;
the plus and minus signs show the in-phase and out-of-phase,
respectively.  
Then the above results imply that a choice of the phase 
at the beginning determines the phases of  the mode-2 and the mode-3, 
while the phase of mode-1 is determined by the boson-fermion coupling.
These results exhibit typical behaviors of the forced vibration
with the external force caused by the boson oscillation,
which is discussed in the following subsection.

\subsection{Modes of the fermion oscillation}
\label{ModeF}

In this subsection we examine the above three modes for 
the fermion oscillations by performing the test simulations
shown in the following.

\begin{figure}[ht]
\includegraphics[scale=0.75]{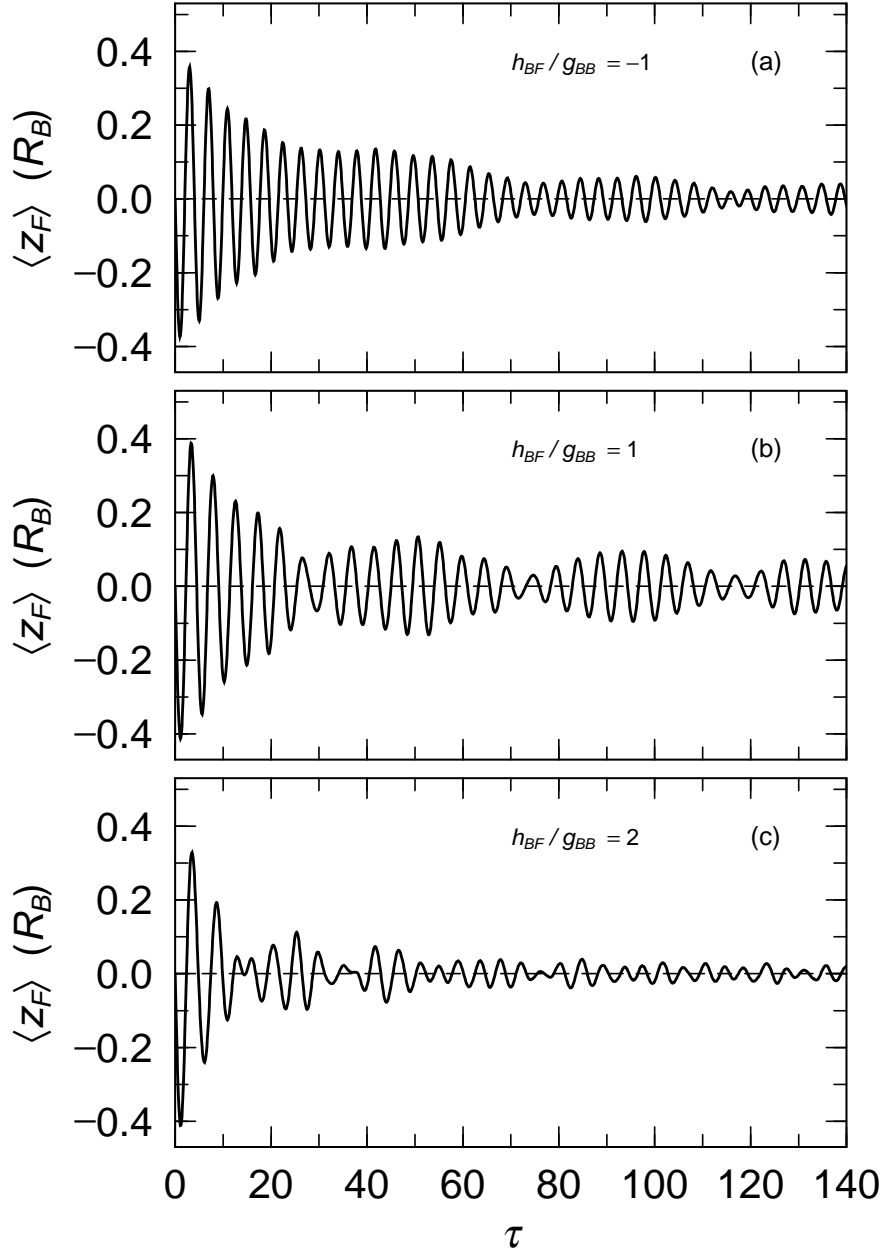}
\caption{\small
Time evolution of $z_F$ (sold lines)
with the boson-fermion coupling $h_{BF} = - g_{BB}$ (a) and
 $h_{BF} = g_{BB}$ (b) when the boson motion is frozen.} 
\label{dpl-FrB}
\end{figure}

First we calculate the fermion dipole oscillations
with the boson motion frozen;
namely the fermions move in the fixed  potential $U_F (\vbr)$ in the 
ground state.
In Fig.~\ref{dpl-FrB} we show the results 
with the boson-fermion coupling
$h_{BF} = - g_{BB}$ (a), $h_{BF} = g_{BB}$ (b) 
and $h_{BF} = 2 g_{BB}$ (c).
In each result we see a beat and damping:
the amplitude of $z_F$ oscillation is about 0.4 in the beginning,
and it becomes about 0.05 after the damping.

\begin{figure}[ht]
\hspace*{-1.0cm}
\includegraphics[scale=0.8]{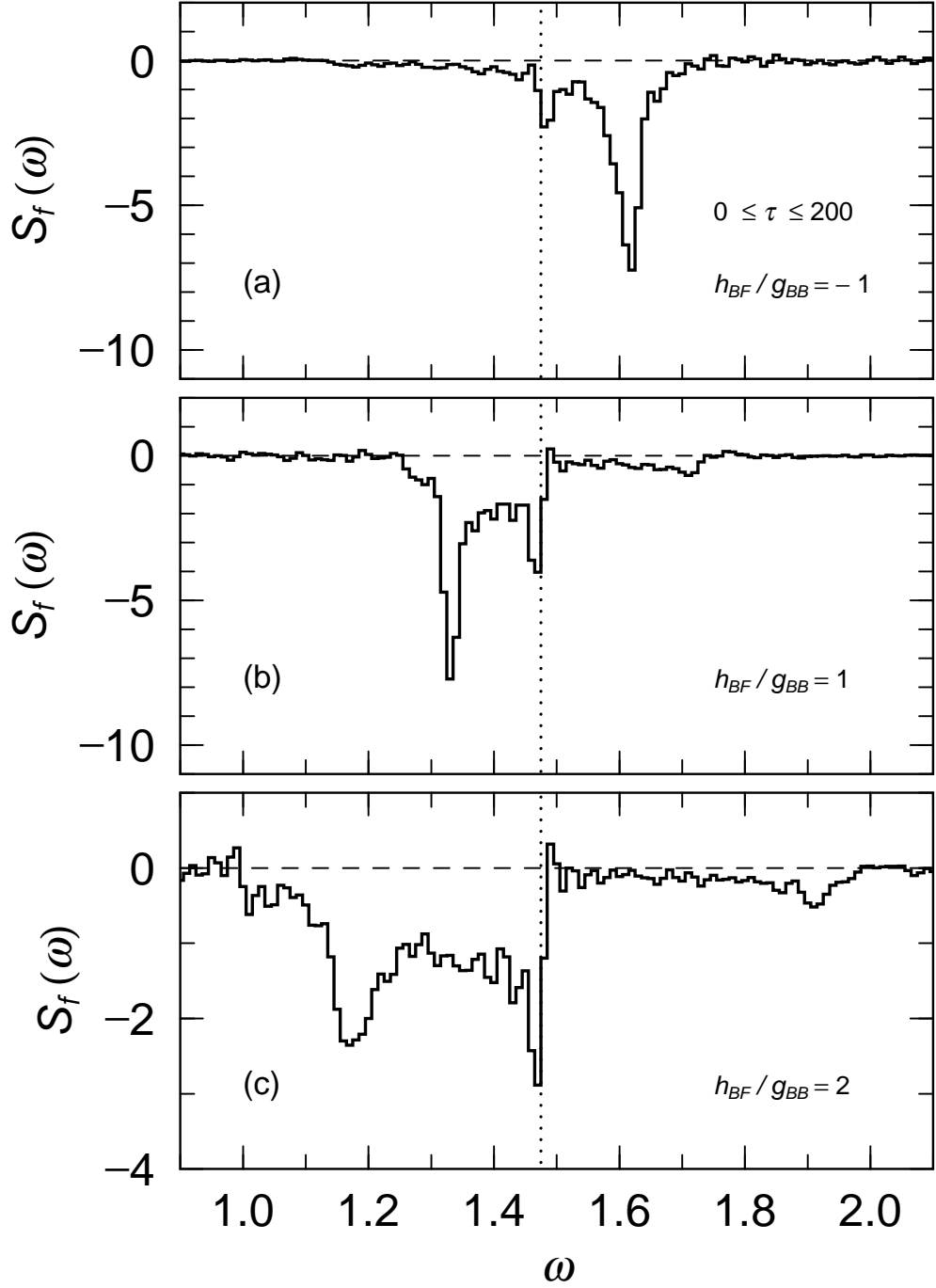}
\caption
{\small 
Strength functions of the fermion oscillation
with $h_{BF}=g_{BB}$ (a), $0.5g_{BB}$ (b), $-0.5g_{BB}$ (c)
and with $h_{BF}=-g_{BB}$ (d)
when the boson motion is frozen.
The dotted lines indicate the fermion trapped frequency $\omega_f$.
}
\label{spcFr}
\end{figure}

In Fig.~\ref{spcFr} 
we plot the strength functions of these fermion oscillations 
with the boson-fermion coupling 
$h_{BF} = -g_{BB}$ (a), $h_{BF} = g_{BB}$ (b) and $h_{BF} = 2g_{BB}$ (c).
In all the results there are two peaks  at $\omega=\omega_f$
and $\omega=\omega_D^f$, but no peak at $\omega=\omega_D^b$.

These results demonstrate that 
mode-2 and mode-3 are intrinsic modes for the fermion oscillation,
and that the mode-1  is caused by the boson oscillation.
A mixture of mode-2 and mode-3 arises the beat,
and this mixing and their widths make the damping 
in the fermion oscillations. 

\begin{figure}[ht]
\hspace*{-1.0cm}
\includegraphics[scale=0.75]{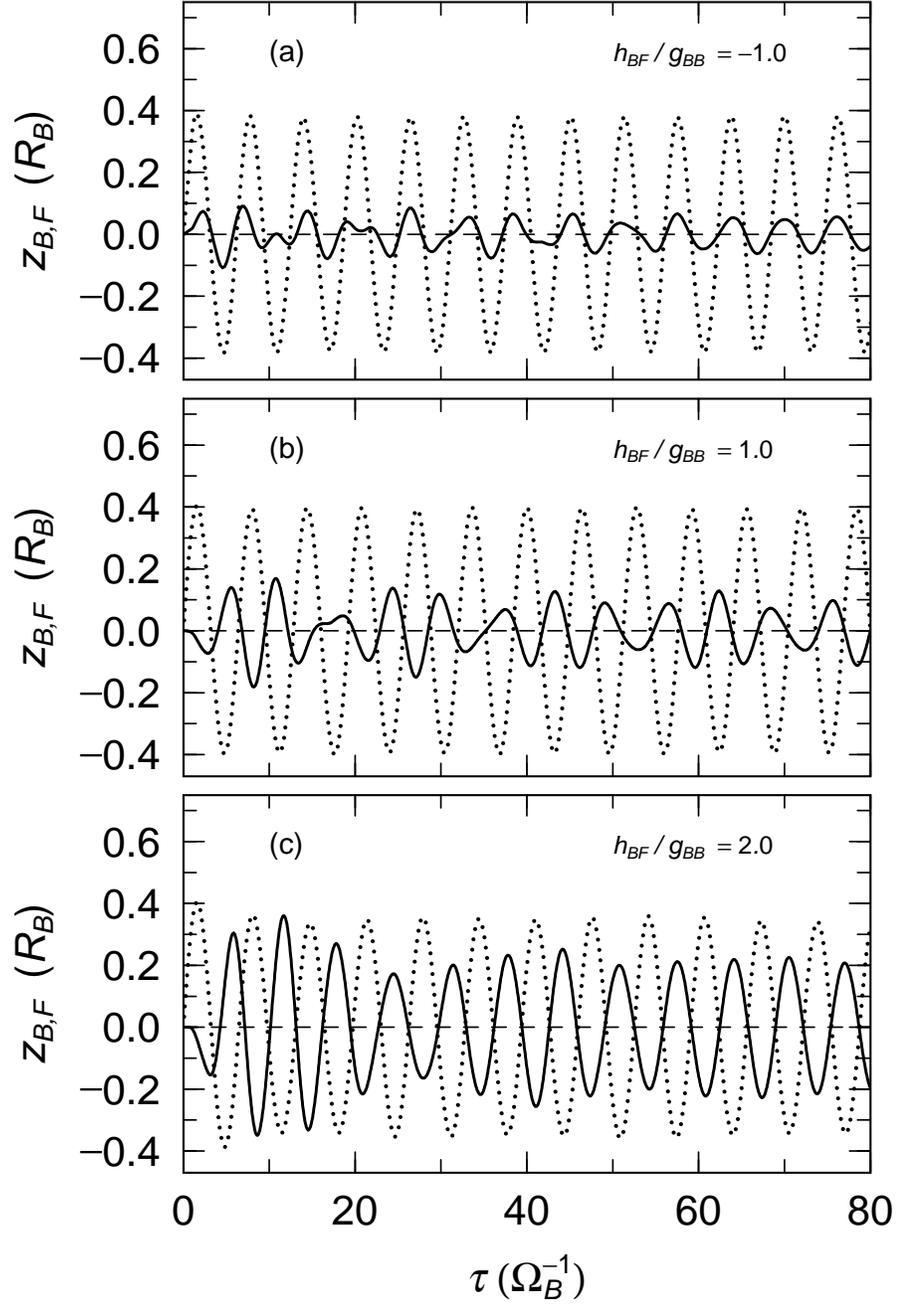}
\caption{\small
Time evolution of $z_B$ (dotted lines) and  $z_F$ (sold lines)
without the boson boost at the beginning
with the boson-fermion coupling $h_{BF} = - g_{BB}$ (a),
 $h_{BF} = g_{BB}$ (b) and  $h_{BF} = 2 g_{BB}$ (c).} 
\label{dplBs}
\end{figure}

\begin{figure}[ht]
\hspace*{-1.0cm}
\includegraphics[scale=0.75,angle=270]{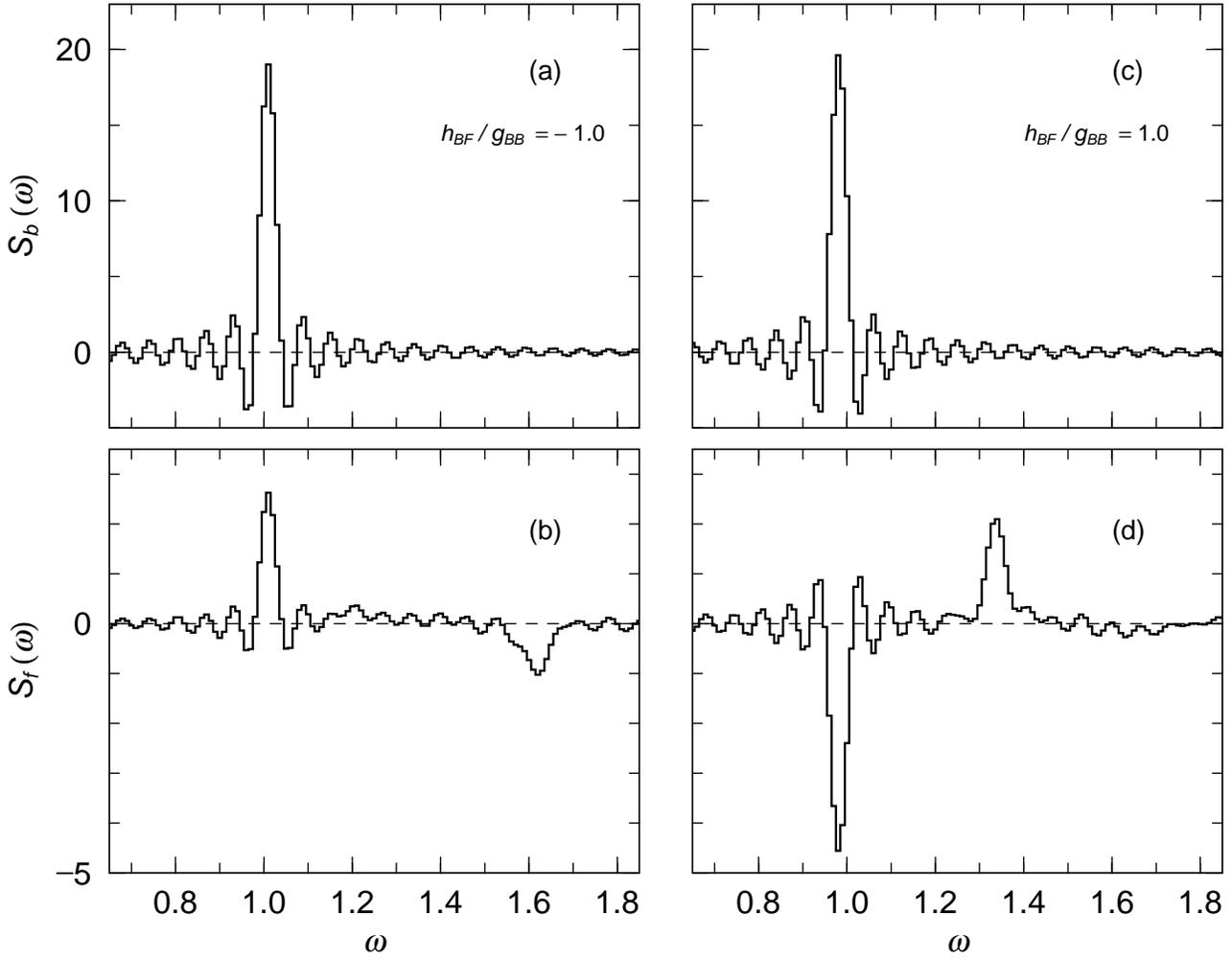}
\caption
{\small 
Strength functions of the boson (a,c) and fermion (c,d) oscillations 
with $h_{BF}= -g_{BB}$ (a,b) and with $h_{BF}= g_{BB}$ (c,d).
The details are shown in Fig.~\ref{dplBs}.
}
\label{spcBs}
\end{figure}

Next we simulate the dipole oscillation
with the initial condition $\lambda_B=0.4$ and $\lambda_F=0$.
In Fig.~\ref{dplBs} 
we show the time-dependences of $z_B$ and $z_F$
with the boson-fermion coupling $h_{BF} = - g_{BB}$ (a), 
$h_{BF} = g_{BB}$ (b) and  $h_{BF} = 2 g_{BB}$ (c).
In Fig.~\ref{spcBs}, furthermore, we plot the strength functions
of the boson oscillation with  $h_{BF} = - g_{BB}$ (a) and 
with $h_{BF} = g_{BB}$ (b)
and those of the fermion oscillation with $h_{BF} = - g_{BB}$ (c) 
and with $h_{BF} = g_{BB}$ (d).
In the fermion strength functions (c,d) there are two peaks 
at $\omega = \omega_D^b$ (mode-1) and $\omega = \omega_D^f$ (mode-3),
and no peak at $\omega = \omega_f$ (mode-2).
In this initial condition, $\lambda_F=0$, the boson oscillation is
a trigger of the fermion oscillation and moves only fermions
inside of the boson occupation region.
Hence these results demonstrate that mode-2 and mode-3
are the dipole motions contributed from 
the fermions outside and inside of the boson occupation region,
respectively. 

The condensed boson density distributes in a smaller region 
than the fermion one,
so that the fermion potential, $U_F$, is separated into two regions,
the inside and outside of the boson occupation region.
Outside  the boson occupation region
the fermion potential, $U_F$, has a simple harmonic oscillator shape 
with the trapped frequency $\omega_f$.  
Inside the boson occupation region, furthermore,
the fermion potential around the central region can also 
be approximately described as
a harmonic oscillator potential with a different trapped frequency.

Thus this fermion motion can be described with two kinds of fluids 
corresponding to the above two modes.
One fluid moves outside the boson occupation region,
and makes a dipole oscillation mode with the frequency, 
$\omega = \omega_f$.
The other fluid moving in the inside region 
makes the dipole oscillation with $\omega = \omega_f^D$.
However the fermion potential does not have a simple harmonic
oscillator shape in the boson occupation region, 
and then their dipole motions, particularly for the latter fluid, are 
not harmonic, and  the oscillation amplitudes decrease.

Now we examine the above consideration in quantum calculations.
We calculate the excited states in RPA and 
compare the results with those in our time-dependent approach.
For this purpose we get the fermion wave functions in the ground states 
in the Hartree-Fock (HF) approximation and solve the RPA equation 
 in the way of Ref.\cite{sogo}.
In the HF calculations the fermion number, $N_f$, must be dependent on
the subshell closure of fermion single particle states, and 
changed from $N_f=1000$;
for example $N_f = 989$ with $h_{BF} = - g_{BB}$,
but this slight change hardly affects the final results.
We use seventeen particle-hole states for bosons and  about five hundred
ones for fermions.

\begin{figure}[ht]
\hspace*{-1.0cm}
\includegraphics[scale=0.8]{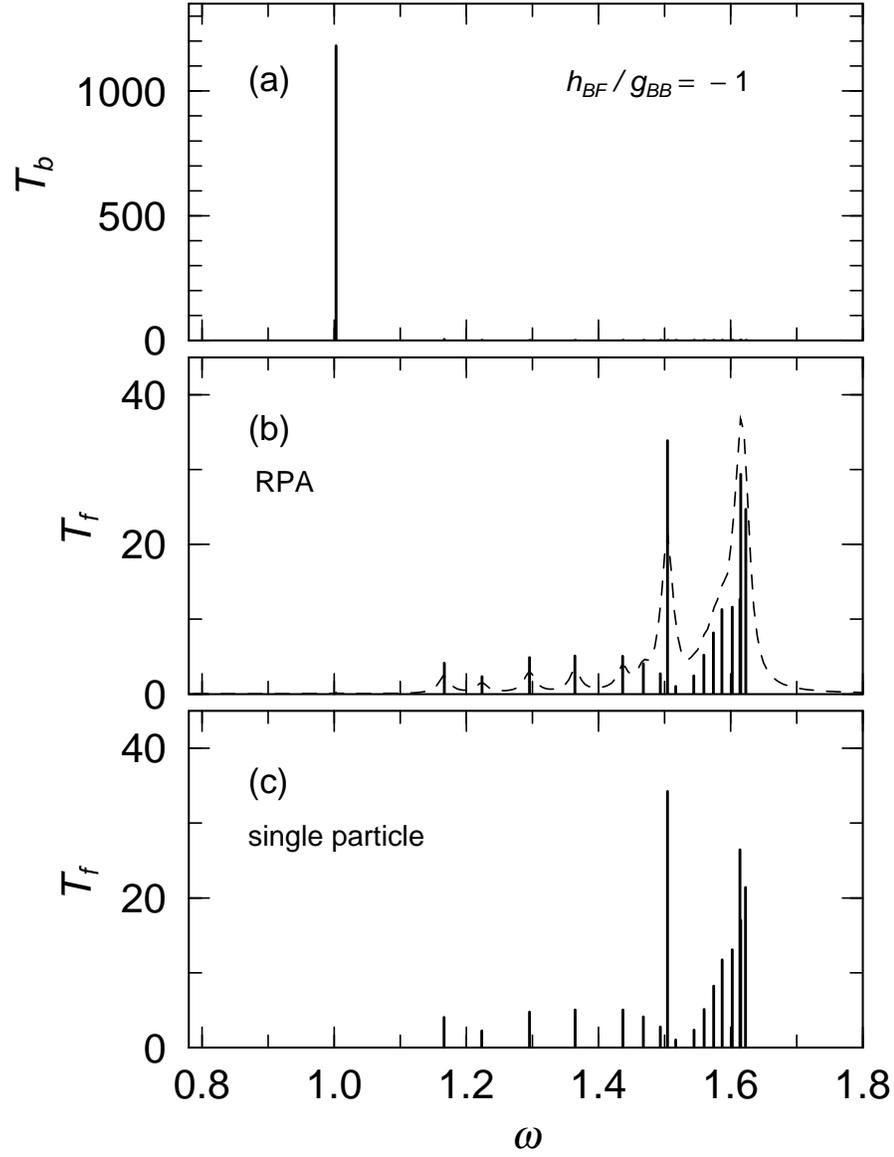}
\caption
{\small 
Transition strengths  versus the excitation energy for the boson (a) 
and  fermion (b) oscillations  with $h_{BF}= -g_{BB}$ in RPA.
The fermion transition strength in the single particle process are
 also ploted in the bottom panel (c).
}
\label{RPAm1}
\end{figure}

Then  we calculate the transition amplitudes from an excited state
$|\Phi_n>$ with excitation energy $\omega_n$ to the ground state
$|\Phi_0>$ for bosons, $A_B$, and for fermions, $A_F$, 
which are written as
\begin{eqnarray}
A_{B} (\omega_n) &=& <\Phi_n| {\hat Z}_B |\Phi_0> ,
\\
A_{F} (\omega_n) &=& <\Phi_n| {\hat Z}_F |\Phi_0>
\end{eqnarray}
with
\begin{eqnarray}
{\hat Z}_B &=&  \int d^3 r \phi^*(\vbr) z \phi(\vbr) ,
\\
{\hat Z}_F &=&  \int d^3 r \psi^*(\vbr) z \psi(\vbr) .
\end{eqnarray}
In Fig.~\ref{RPAm1} we show the boson 
$T_B(\omega_n) = |A_B(\omega_n)|^2/4\pi$ (a) and 
fermion transition strengths 
$T_F(\omega_n) = |A_F(\omega_n)|^2/4\pi$ (b) 
as functions of the excitation energy $\omega$ with $h_{BF} = - g_{BB}$.

In order to visualize the distribution for the fermion transition strength,
we introduce an artificial width, $\Gamma$, and draw the curve of $T_F^s(\omega)$
defined by
\begin{equation}
T_F^s (\omega) = \frac{1}{\pi} \sum_n T_F (\omega_n) 
\frac{\Gamma}{\Gamma^2 + (\omega - \omega_n)^2} . 
\label{lorsm}
\end{equation} 
In our calculation we take $\Gamma = 0.01$, and plot 
$0.018 \times T_F^s(\omega)$
in the second panel (b), where the arbitrary factor, 0.018, is
introduced to plot the curve with the same scale of $T_F$ in the same figure.

In these results there is a clear peak at $\omega=1.0$ 
in the boson transition and two clear peaks at $\omega = 1.50$ and
$1.61$ in the fermion transition.
These two peak positions of $T_F$ are close to 
the frequency of the mode-2, $\omega_f$, and that of the mode-3, $\omega_D^f$ 
in our time-dependent approach.
Furthermore we see several modes with the excitation energy between
 $\omega_f$ and $\omega_D^f$.

We would like to comment that
there is a peak $T_F = 0.075$ at $\omega \approx 1.0$, but 
it is too small to be visible.
In the RPA calculation, thus, there is not any any strong co-moving
mode, and hence   
the fermion transitions must be almost explained as single particle processes.
In Fig.~\ref{RPAm1}c we plot the fermion transition
strength, $T_F$, in the single particle process;
the results of Figs.~\ref{RPAm1}b and \ref{RPAm1}c are almost the same.

\bigskip

Hence the dipole oscillation behavior can be explained as follows.
The condensed bosons occupy one single particle state, and
the boson oscillation has one mode and does not damp.
In contrast the fermions occupy many single particle states,
and the fermion oscillation includes two intrinsic modes
which are contributed from the fermi gases outside and inside
of the boson occupation region, respectively.
Since the fermion potential cannot be critically separated,
there are several modes with frequencies between $\omega_f$ and $\omega_D^f$.
Because of this statistical difference, furthermore,
the boson oscillation almost one-sidedly affects 
the fermion oscillation.
Then the fermion oscillation has one more mode
 caused by the external vibration force which is the boson oscillation.
As a result the fermi gas gradually loses its intrinsic modes 
and finally oscillates with the same frequency of the condensed 
bose gas.

\subsection{Comparison with the Sum-Rule Approach}

In this subsection we discuss the boson-fermion coupling dependence
of the oscillation frequencies by comparing our results
with the sum-rule approach \cite{miyakawa}.

In Fig.~\ref{spcBs} 
we show the intrinsic frequencies of boson and fermion 
with full diamonds and circles, respectively.
For references  the fermion frequencies with the boson motion frozen 
are plotted with the open circles which are connected with the long-dashed line,
and the trapped frequency of the fermion with the dotted line.
In addition we give the results of the RPA, which is defined as the peak
energy of $T_F^s$, with the asterisk.
Our results about the fermion oscillation in the full calculation
(full circles) agree with those in the boson motion frozen (open
circles) and those in RPA (asterisk).

The sum-rule approach \cite{miyakawa} gives
the intrinsic frequencies of the boson and fermion 
dipole oscillations as
\begin{eqnarray}
\omega_D^b &=& \sqrt{1 - \frac{1}{N_b}V_{DP}} ,
\label{sumB}
\\
\omega_D^f &=& \sqrt{\omega_f^2 - \frac{1}{m_f N_f} V_{DP}}
\label{sumF}
\end{eqnarray}
with
\begin{equation}
V_{DP} = h_{BF} \int d^3 r 
\frac{\partial \rho_B}{\partial z} \frac{\partial \rho_F}{\partial z} .
\label{Vsum}
\end{equation}
In the same figure we also plot the results of the sum-rule approach 
with the thick dashed lines.
The above results in the sum-rule approach agree with our results 
when $h_{BF} < 0$, though
the two results show discrepancy when  $h_{BF} > 0 $.

It is known from Eqs.(\ref{sumB}), (\ref{sumF}) and (\ref{Vsum}) that 
the frequency calculated with the sum-rule is sensitive to 
the density distribution.
As discussed in Sec.III-A the derivative of the fermion density 
with respect to the radial coordinate in the boson occupation region
is negative, $\partial \rho_F/\partial r < 0$, 
when $h_{BF} \lesssim g_{BB}$, and  
positive, $\partial \rho_F/\partial r > 0$,  when $h_{BF} \gtrsim g_{BB}$, 
while the derivative of the boson density, 
$\partial \rho_B / \partial r < 0$, in the both coupling region.
As the boson-fermion coupling, $h_{BF}$,  increases, hence, 
the fermion frequency in the sum-rule approach decreases when $h_{BF}<0$, 
is not varied when $0<h_{BF}<1$, and increases when $h_{BF}>0$ .

The sum-rule approach is available only when the transition strength is
concentrated to one excited state.
In BF mixtures, however, the fermion transition strength distributes
mainly to the two modes, and then the sum-rule approach cannot predict
the correct intrinsic frequency of the fermion oscillations.

Both our time-dependent approach and the RPA calculation explain
the two modes in the fermion oscillation.
However the RPA can be available only in minimal oscillations,
and we should  examine the time dependence of $z_{B,F}$ in RPA.

Here we calculate the time evolutions $z_B$ and $z_F$ in RPA.
By using the initial boost given in Eqs.(\ref{bin}) and (\ref{fin})
with the perturbative way,
the initial state in the dipole oscillation becomes
\begin{eqnarray}
|\Phi(\tau=0)> &=&
e^{i \left\{ \lambda_B {\hat Z}_b + \lambda_F {\hat Z}_f \right\}} |\Phi_0>
\nonumber \\
&\approx& |\Phi_0> +
i \left\{ \lambda_B {\hat Z}_b + \lambda_F {\hat Z}_f \right\} |\Phi_0> .
\nonumber \\
&=&  |\Phi_0> + i \sum_n \left\{ \lambda_B A_B(\omega_n)
+ \lambda_F A_F (\omega_n) \right\} |\Phi_n> .
\end{eqnarray}
This initial boost is equivalent to that given in Eqs.(\ref{bin}) and
(\ref{fin}) in the first order of $\lambda_{B,F}$.
Then, the time-dependence of $z_B$ and $z_F$  are given as
\begin{eqnarray}
z_B(\tau) &=& \frac{1}{N_b} \sum_n <\Phi(\tau)|{\hat Z}_b|\Phi(\tau)>
\nonumber \\
&=&
\frac{2}{N_b} \sum_n A_B(\omega_n)
(\lambda_B A_B(\omega_n) + \lambda_F A_F(\omega_n) ) \sin(\omega_n \tau) ,
\label{zBrpa}
\\
z_F (\tau) &=&  \frac{1}{N_f} \sum_n <\Phi(\tau)|{\hat Z}_f|\Phi(\tau)>
\nonumber \\
&=&
\frac{2}{N_f} \sum_n A_F(\omega_n)
(\lambda_B A_B(\omega_n) + \lambda_F A_F(\omega_n) ) \sin(\omega_n \tau) .
\label{zFrpa}
\end{eqnarray}

\begin{figure}[ht]
\hspace*{-1.0cm}
\includegraphics[scale=0.8]{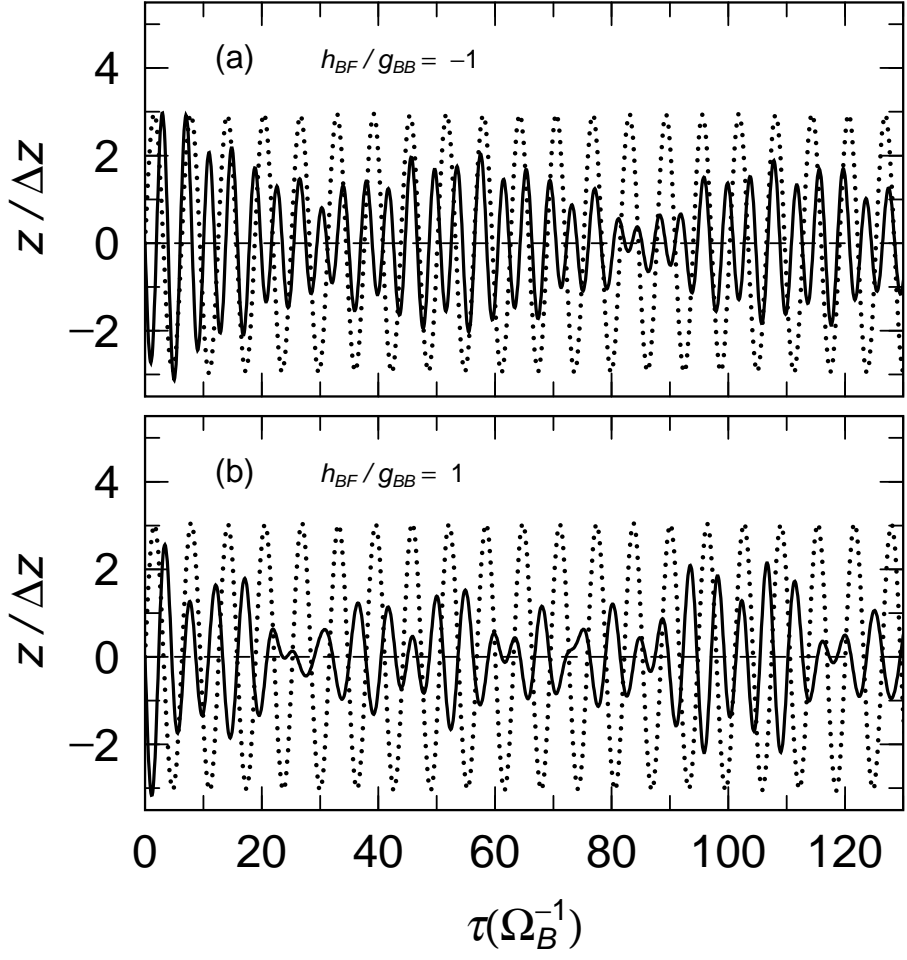}
\caption
{\small 
Time evolutions of $z_B$ (solid line) and $z_F$ (dashed line) 
in RPA with $h_{BF}/g_{BB} = -1$ (a) and  $h_{BF}/g_{BB} =  1$ (b).
}
\label{tevRPA}
\end{figure}

In Fig.~\ref{tevRPA}  we show the results for boson (dotted line) 
and fermions (solid line)
in the out-of-phase oscillations with $h_{BF}=-g_{BB}$ (a)
and  $h_{BF}=g_{BB}$ (b) by using
the condition $\lambda_B = - \lambda_F = \Delta z$. 
The amplitudes of the oscillations are scaled by
$\Delta z$.
Comparing these results with those in 
the time-dependent approach (Fig.~\ref{dplRbK}),  
we can see a clearer beat and weaker damping in the fermion oscillations, 
and the period of the
fermion oscillation dose not become the same as that of the boson
oscillation even in the later time stage.
The boson oscillation does not affect the fermion osciilation in RPA 
as strongly as in the time dependent approach.

\begin{figure}
\includegraphics[scale=0.7,angle=270]{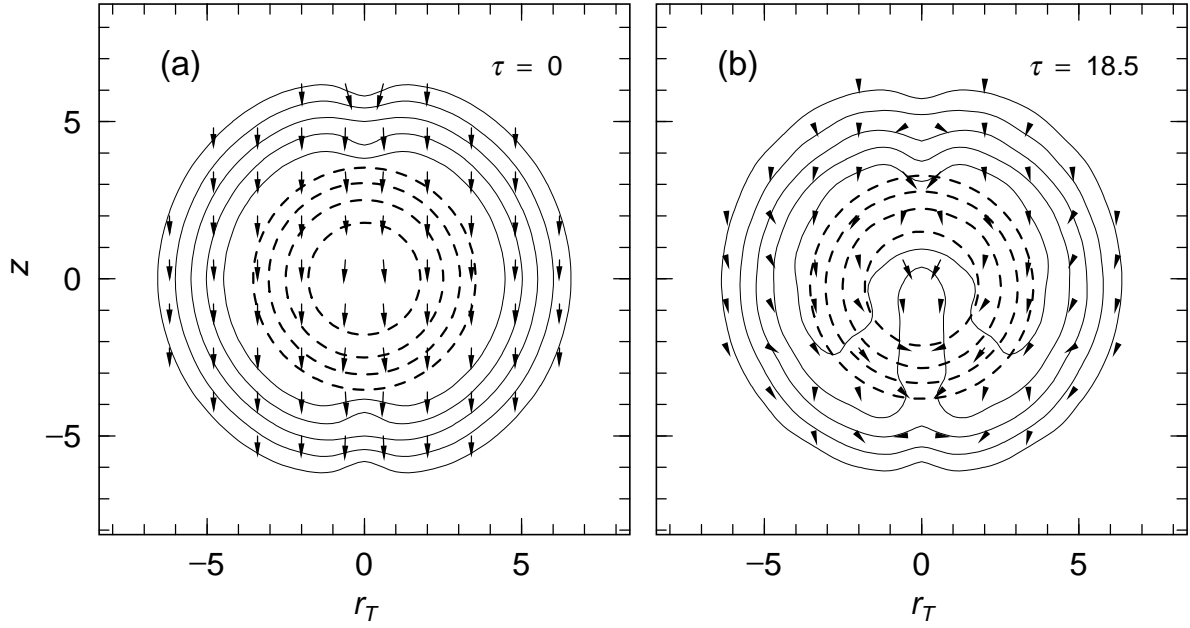}
\medskip
\caption{\small
Density distribution and velocity fields.
at $\tau =0$ (a) and $\tau=18.5$ (b).
The dashed and solid contour lines  
represent the density distributions of the bose and fermi
gases, respectively,
and the arrows indicate the velocities of the fermi gas.
}
\label{velFr}
\end{figure}

In order to inspect the reason for this difference more, 
we show the density distribution
and the velocity fields  at $\tau = 0$ (a) and at $\tau = 18.5$ (b) 
in Fig.~\ref{velFr}.
The dashed and solid contour lines represent the density distributions
for boson and fermion, respectively, and 
the arrows indicate the fermion velocities.
At both times  the $z$-component of
the CM position of fermi gas is zero, $z_F = 0$.

At $\tau = 0$ the boson and fermi gases move 
with unique velocities (a) in all positions.
At $\tau = 18.5$, however, 
the velocity of the fermi gas is directed along the contour lines 
of the density, and its density becomes dilute in the space region, 
$z<0$ and $|\vbr_T| \lesssim 1$. 
The fermi gas moves to the surface part of the bose gas
because of the repulsive force from the bose gas.
This motion is very natural from the aspect of the hydrodynamics.
 
The RPA approach can describe only minimal oscillations 
with one-particle one-hole excitation energy,
where the density distribution is slightly changed from that at the
ground state.
In the case of $h_{BF}=g_{BB}$ and the boson motion frozen, for example,
its excitation energy is about 80 $\hbar \Omega_B$, and the system
cannot be described with simple one-particle and one-hole states. 
In such non-small amplitude oscillations 
the fermi gas changes its density 
distribution and  moves to the place with low resistance. 
Hence  the actual oscillation 
of the BF mixing gases are not so simple to describe with the RPA
and sum-rule approaches.

\section{Summary}

In this paper we study the collective dipole oscillation in the
BF mixtures by solving the time evolutions of the system directly
with the TDGP and Vlasov equations. 
The calculational results reveal 
that the boson and fermion oscillations
make different behaviors reflected by their  statistics.
The condensed bosons occupy one single particle state, but
the fermions distribute in many single particle states. 
Then the boson oscillation has 
only one mode and does not have damping, but
the fermion collective oscillation includes various modes.
In the BF mixtures, especially, the fermi gas is separated 
into the two regions, inside and outside of the boson occupation region,
and these two fluids oscillate with different periods.

Furthermore the boson oscillation almost one-sidedly affects 
the fermion oscillation, and then
the fermion oscillation includes three modes: 
one is a mode  of the forced vibration caused by the boson oscillation,
the other modes are two fermion intrinsic modes 
contributed from the fermi gases inside and outside
of the boson occupation region, respectively.
In addition there are several small modes in the fermion oscillations
with frequencies between the two intrinsic frequencies.
These modes cause the beat and damping in the fermion oscillation. 
As the result the fermion motion gradually loses the strength 
of its intrinsic modes and finally oscillates 
with the same frequency of the condensed bose gas.

The RPA approach can also explain the frequencies of the boson and fermion 
intrinsic modes.
The qualitative behavior of the transition strength in RPA
oscillations is similar to the strength function in our approach.
Nevertheless the time evolutions of $z_F$ are not the same, particularly 
in the later time stage, because 
the forced oscillation modes, mode-1, does not largely contribute to
the whole oscillation in RPA. 
RPA is available only in  the minimal oscillation, and cannot describe
the change of the density distribution in the time evolution process.
In actual experiments the amplitude is not so small, 
and hence we must solve the time-dependent process 
of the collective oscillations directly.

In this work we do not take into account two body collisions and
thermal boson \cite{JackZar}.
In the system $N_b \gg N_f$ at zero temperature, 
the thermal bosons are very few, and then the two body collisions
do not play significant roles in the dynamical process. 
However, it is not easy to establish
gases at such low temperatures in actual experiments.
In the future we need to introduce the two body collision terms
into our approach  \cite{BUU1,TOMO1}.

\newpage

\newpage

\newpage

\newpage

\newpage

\newpage

\newpage

\newpage

\newpage

\newpage

\newpage

\begin{figure}
\includegraphics[scale=0.7,angle=270]{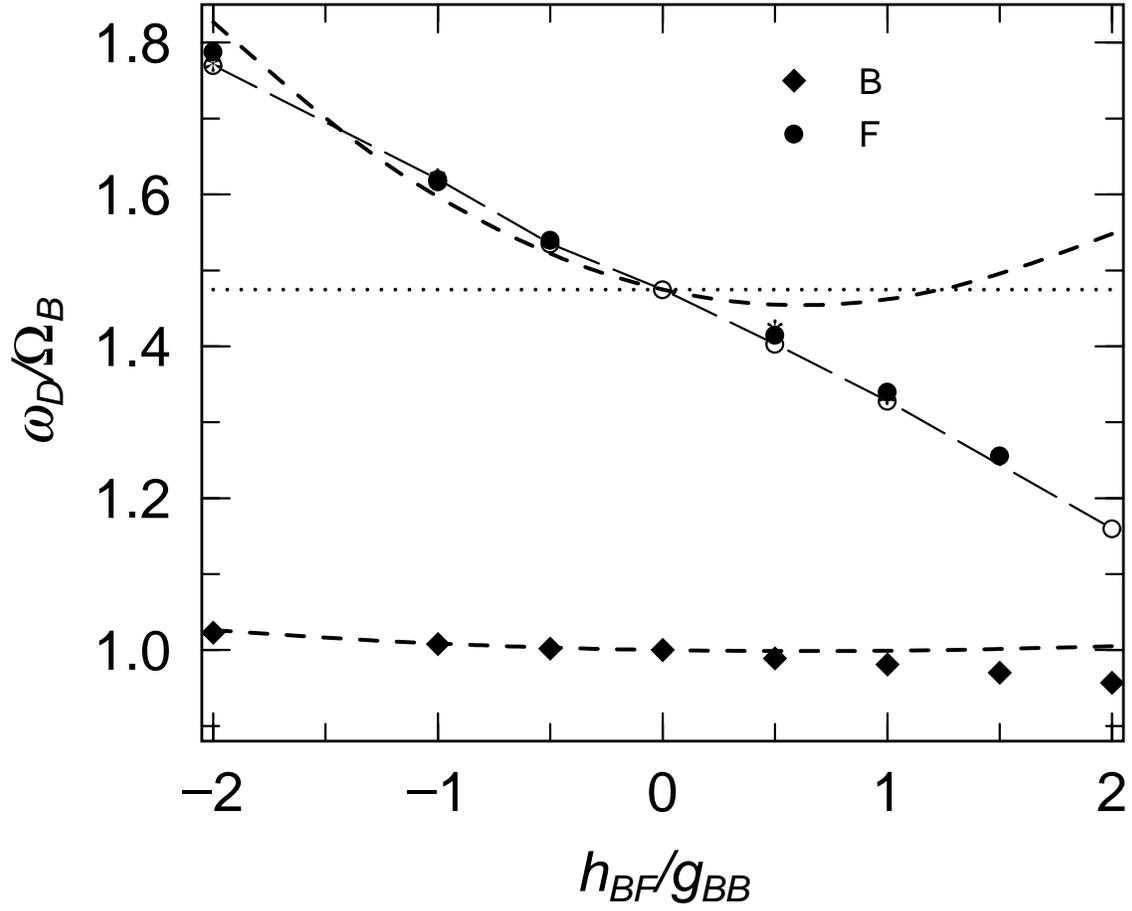}
\medskip
\caption{\small
Intrinsic frequencies of boson and fermion oscillations 
versus the boson-fermion coupling.
Full diamonds and full circles represent 
the intrinsic frequencies of boson and
fermion oscillations, respectively.
The open circles, which are connected with the long-dashed line, 
indicate the fermion intrinsic frequency with the boson motion frozen.
The thick dashed and dotted lines denote the results of the sum-rule
and the trapped frequency of the fermion, respectively.
}
\label{frDP}
\end{figure}

\newpage

\newpage

\end{document}